\documentclass[conference]{IEEEtran}

\usepackage{graphicx,psfrag,epsfig,epsf,latexsym,hhline,amsmath,amssymb,multirow}
\usepackage{pst-plot}
\usepackage{pstricks-add}
\usepackage{pifont}
\usepackage{graphicx}
\usepackage{amsthm}
\usepackage[linesnumbered,ruled,vlined]{algorithm2e}
\usepackage[noadjust]{cite}
\usepackage{array}
\usepackage{blindtext}
\usepackage{etoolbox}
\usepackage{comment}
\usepackage{epstopdf}
\usepackage{hyperref}

\newcommand\xqed[1]{%
  \leavevmode\unskip\penalty9999 \hbox{}\nobreak\hfill
  \quad\hbox{#1}}
\newcommand\demo{\xqed{$\blacksquare$}}

\SetCommentSty{mycommfont}

\SetKwInput{KwInput}{Input}                
\SetKwInput{KwOutput}{Output}              

\newcolumntype{M}[1]{>{\centering\arraybackslash}m{#1}}
\newcolumntype{P}[1]{>{\centering\arraybackslash}p{#1}}

\newcommand{\msf}[1]{\mathsf{#1}}

\newcommand{\SNR}{\msf{SNR}}

\newcommand{\E}{\mathbb{E}}

\newcommand{\iid}{i.\@i.\@d.\ }

\theoremstyle{definition}\newtheorem{lemma}{Lemma}
\theoremstyle{definition}\newtheorem{proposition}[lemma]{Proposition}
\theoremstyle{definition}
\theoremstyle{definition}

\newtheorem{remark}{Remark}

\begin{document}
\title{Downlink Transmission under Heterogeneous Blocklength Constraints: Discrete Signaling with Single-User Decoding}
\author{ Min Qiu$^\dag$, Yu-Chih Huang$^*$, and Jinhong Yuan$^\dag$\\
$^\dag$School of Electrical Engineering and Telecommunications, University of New South Wales, Sydney, Australia\\
$^*$Institute of Communications Engineering, National Yang Ming Chiao Tung University, Hsinchu City, Taiwan\\
E-mail: \{min.qiu, j.yuan\}@unsw.edu.au, jerryhuang@nycu.edu.tw  \\


}

\maketitle

\begin{abstract}
In this paper, we consider the downlink broadcast channel under heterogenous blocklength constraints, where each user experiences different interference statistics across its received symbols. Different from the homogeneous blocklength case, the strong users with short blocklength transmitted symbol blocks usually cannot wait to receive the entire transmission frame and perform successive interference cancellation (SIC) owing to their stringent latency requirements. Even if SIC is feasible, it may not be perfect under finite blocklength constraints. To cope with the heterogeneity in latency and reliability requirements, we propose a practical downlink transmission scheme with \emph{discrete signaling} and \emph{single-user decoding, i.e., without SIC}. In addition, we derive the finite blocklength achievable rate and use it for guiding the design of channel coding and modulations. Both achievable rate and error probability simulation show that the proposed scheme can operate close to the benchmark scheme which assumes capacity-achieving signaling and perfect SIC.


\end{abstract}

%

\section{Introduction}
Future generation wireless systems are expected to support a wide range of smart devices while achieving high spectrum and energy efficiency. It is known that the conventional orthogonal multiple access (OMA) is difficult to meet these future requirements due to inefficient use of radio resources \cite{tse_book}. This calls for more efficient multiplexing schemes for providing connectivity to the growing number of devices. In fact, this call may have been partially answered in the classical studies on the class of broadcast channel (BC) \cite[Ch. 6.2.2]{tse_book}, where superposition coding and successive interference cancellation (SIC) are two key ingredients for achieving the capacity of the scalar Gaussian BC effectively. Building upon this result, many popular multiple access schemes have successfully adopted these two techniques for enabling simultaneous wireless access for multiple users/devices. These schemes are sometimes referred to as non-orthogonal multiple access (NOMA) \cite{Ding17J,9693417}. Since the major development of finite blocklength information theory \cite{5452208}, refining the performance of multiple access schemes with homogeneous and finite blocklength constraints has drawn some attention, e.g., \cite{8277977,8345745,8933345}.


Recently, the coexistence between enhanced mobile broadband (eMBB) and ultra-reliable low-latency communications (URLLC) was discussed in \cite{8403963,8476595,8647460}. Notably, \cite{8476595,8647460} introduce heterogenous NOMA to serve both eMBB and URLLC users using the same time/frequency resources. However, in the downlink BC, the decoding of URLLC signals cannot leverage SIC but can only treat the \emph{partially} received eMBB symbols as noise owing to the URLLC latency constraint. To address this issue, \cite{9518265,9838392} introduces early decoding which allows the strong user with short packets to perform SIC based on partially received superimposed symbol blocks when certain conditions regarding the channel and blocklength are met. That said, SIC could introduce error propagation, extra decoding latency and complexity and may compromise users' privacy. In addition, most of the works assume using capacity-achieving signaling such as Gaussian codes \cite[Eq. (6)]{7605463} and shell codes, i.e., codewords drawn from a power shell \cite[Eq. (5)]{7605463}. In practice, the current prevailing approach is to adopt channel coding with discrete constellations, e.g., quadrature amplitude modulation (QAM) \cite{TS138212_v16p8}. This motivates us to design practical schemes based on \emph{discrete signaling} and \emph{single-user decoding (SUD)} for multiplexing heterogeneous services, aiming at approaching the performance under Gaussian codes or shell codes with perfect SIC.


Discrete signaling and treating interference as noise (TIN), i.e., SUD, in the infinite blocklength regime have been investigated in our previous works \cite{8291591,9535131}. However, under \emph{heterogeneous finite blocklength} and \emph{non-vanishing error probability} constraints, how to effectively manage heterogeneous interference across received symbol sequences has not been investigated. Moreover, the characterization of second-order achievable rates with practical modulations and TIN for this scenario is lacking. In this paper, we introduce a new downlink transmission scheme based on practical discrete signaling and low-complexity SUD for the downlink BC with heterogeneous blocklength and error probability constraints. We first design the discrete input distribution for each downlink user according to the heterogeneity in interference statistics. We then analyze the achievable rate of each user given its own blocklength and error probability requirements. Simulation results show that in terms of the achievable rate and error probability performance, the proposed scheme with QAM and TIN can operate close to the benchmark scheme which assumes using capacity-achieving signaling and perfect SIC.



\textbf{Notations}: All logarithms are base 2. $Q^{-1}(x)$ denotes the inverse of Q function $Q(x)=\int^{\infty}_{x}\frac{1}{\sqrt{2\pi}}e^{-\frac{t^2}{2}}dt$. We write $f(x) = O(g(x))$ if $\exists M\in\mathbb{R}^+,x_0\in\mathbb{R}$ such that $|f(x)| \leq Mg(x), \forall x\geq x_0$. Random variables are represented by uppercase letters, e.g., $X$, and their realizations are represented by lowercase letters, e.g., $x$. $X^{[n]}$ denotes the sequence $X[1],\ldots,X[n]$.


\section{System Model}\label{sec:model}
We consider a scalar downlink BC that consists of one transmitter and two receivers. We leave the generalization to the $K$-user case in our journal version \cite{JSACQiu22}. We denote by $\boldsymbol{x}_i \in \mathbb{C}^{N_i}$ the transmitted packet of coded symbols for user $i$, where $i\in \{1,2\}$ and $N_i$ is the length. We assume that $N_1 \leq N_2$ without loss of generality. Thus, the proposed scheme do not preclude the homogeneous blocklength case. The transmitter broadcasts the superimposed coded symbols of length $N_2$
\begin{align}
\boldsymbol{x} = ([\boldsymbol{x}_1,\boldsymbol{0}_{N_2-N_1}]+\boldsymbol{x}_2)=\boldsymbol{x}'_1+\boldsymbol{x}'_2 \in \mathbb{C}^{N_2}, \label{eq:sys1}
\end{align}
to both users, where we define $\boldsymbol{x}'_i \triangleq [\boldsymbol{x}_i,\boldsymbol{0}^{N_2-N_i}]$. Here, we consider that $\boldsymbol{x}_1$ needs to be transmitted as soon as possible owing to its urgency, e.g., URLLC services. Thus, $\boldsymbol{x}_1$ is superimposed with the first $N_1$ symbols of $\boldsymbol{x}_2$. We emphasize that the position of $\boldsymbol{x}_1$ in $\boldsymbol{x}'_1$ does not affect our scheme. In addition, we have the following individual power constraint $P_i$ and total power constraint $P$, and $P_i \leq P$.
\begin{align}
&\frac{1}{N_i}\sum\nolimits_{j=1}^{N_i}|x_i[j]|^2 \leq P_i,\label{eq:power_ind}\\
&\frac{1}{N_2}\sum\nolimits^{N_2}_{j=1}(|x'_1[j] |^2+|x'_2[j] |^2) \leq P.\label{eq:power_total}
\end{align}


We denote by $h_i \in \mathbb{C}$ the channel of user $i$, $i\in\{1,2\}$. We assume that $h_i$ is subject to quasi-static fading, i.e., $h_i$ remains unchanged over the duration of each transmission frame. The received signals are given by
\begin{align}
y_1[j] =& h_1(x_1[j]+x_2[j])+z_1[j], \;j=1,\ldots,N_1,\label{eq:y1} \\
y_2[j] =& \left\{ {\begin{array}{*{20}{c}}
h_2(x_1[j]+x_2[j])+z_2[j],\; j=1,\ldots,N_1\\
h_2x_2[j]+z_2[j], \;j=N_1+1,\ldots,N_2\\
\end{array}} \right.\label{eq:y2},
\end{align}
respectively, where $z_i[j] \sim \mathcal{CN}(0,1)$ is the i.i.d. Gaussian noise. Clearly, user 2's symbols $y_2[N_1+1],\ldots,y_2[N_2]$ are interference-free. We assume that the transmitter has the knowledge of channel magnitudes while the receiver has full channel state information. We stress that even for such a fundamental channel model, many problems are yet to be solved, e.g., the optimal communication strategy, the optimal input distribution, and the second-order converse. Two important performance metrics are jointly considered in this paper, namely the achievable rate $R_i$ and upper bound on the average decoding error probability $\epsilon_i$ for user $i\in\{1,2\}$.

\section{Proposed Discrete Signaling with SUD}
In this section, we introduce the proposed scheme with discrete signaling and TIN. Although we use binary codes and QAM as the underlying channel codes and constellations, respectively, our scheme does not preclude the use of non-binary codes \cite{8066336} and multi-dimensional constellations \cite{8291591}.

We assume that $|h_1|>|h_2|$. This corresponds to the interesting case where the URLLC user, i.e., user 1, is the strong user but performing SIC may not be feasible based on partially received superimposed symbols. The case of $|h_1|<|h_2|$ will be discussed later.

\subsubsection{Encoding}
For user $i\in\{1,2\}$, a length-$k_i$ binary source sequence $\boldsymbol{u}_i$ is encoded into a length-$n_i$ binary codeword $\boldsymbol{c}_i$. Codeword $\boldsymbol{c}_i$ is then interleaved, i.e., using bit-interleaved coded modulations (BICM) \cite{CIT-019}, and modulated onto a length-$N_i$ sequence $\boldsymbol{v}_i$. Each user only uses a \emph{single} channel. This ensures that the encoding and decoding (TIN) complexities for each user are the same as in the single-user case.

\subsubsection{Modulation Mapping}\label{sec:2u_const}
To handle heterogeneous interference as shown in \eqref{eq:y2}, user 2 uses two sets of constellations $\Lambda_{2,1}$, and $\Lambda_{2,2}$ while user 1 users one constellation set $\Lambda_1$. Specifically, the modulated symbols satisfy $v_1[j] \in \Lambda_1, \forall j \in \{1,\ldots,N_1\}$ for user 1 and $v_2[j] \in \Lambda_{2,1}, \forall j \in \{1,\ldots,N_1\}$, $v_2[j] \in \Lambda_{2,2}, \forall j \in \{N_1+1\ldots,N_2\}$ for user 2. Let $\Lambda_1$, $\Lambda_{2,1}$, and $\Lambda_{2,2}$ represent three regular QAM constellations with zero means and minimum distance 1. We further define $m_1\triangleq \log |\Lambda_1|$, $m_{2,1}\triangleq \log |\Lambda_{2,1}|$ and $m_{2,2}\triangleq \log |\Lambda_{2,2}|$ to be the modulation orders. The relationship between codeword length $n_i$ and coded symbol length $N_i$ for user $i$ satisfies
\begin{align}
n_1 =& N_1m_1, \label{u1n1}\\
n_2 =& N_1m_{2,1}+ (N_2-N_1)m_{2,2}.\label{u2n2}
\end{align}

We then introduce the design criteria for modulation orders. To do so, we first need to introduce the following sub-block power constraints. Since user 2 uses two constellation sets, we can decompose user 2's symbol block $\boldsymbol{x}_2$ into two sub-blocks. The power constraints for these sub-blocks satisfy
\begin{align}
&\frac{1}{N_i-N_{i-1}}\sum\nolimits_{j=N_{i-1}+1}^{N_i}|x_2[j]|^2 \leq P_{2,i},i\in\{1,2\},\label{eq:power_c2sub}\\
&\frac{N_1}{N_2}P_{2,1}+\frac{N_2-N_1}{N_2}P_{2,2}  = P_2,\label{eq:power_c2p}\\
&\frac{N_1}{N_2}(P_1+P_{2,1})+\frac{N_2-N_1}{N_2}P_{2,2}= P\label{eq:power_c2c1p},
\end{align}
where \eqref{eq:power_c2sub} gives the sub-block power constraint for $\boldsymbol{x}_2$ and $N_0=0$, \eqref{eq:power_c2p} gives the relationship between sub-block power constraint \eqref{eq:power_c2sub} and individual power constraint \eqref{eq:power_ind} for user 2, and \eqref{eq:power_c2c1p} gives the relationship between each user's power constraint and total power constraint \eqref{eq:power_total}. The introduction of sub-block power constraints allows to assign different power to each sub-block of $\boldsymbol{x}_2$ for handling heterogeneous interference. Then, we introduce the following constraints on modulation orders $m_1$, $m_{2,1}$, and $m_{2,2}$
\begin{align}
m_1\hspace{-1mm}+\hspace{-1mm}m_{2,1} \leq& \left\lfloor\log  \left(1\hspace{-1mm}+\hspace{-1mm}6(P_1\hspace{-1mm}+\hspace{-1mm}P_{2,1})\max\{|h_1|^2,|h_2|^2\}\right)\right\rfloor,\label{con1}\\
m_{2,1}\leq& \left\lfloor\log  \left(6(P_1+P_{2,1})|h_2|^2\right)\right\rfloor,\label{con2}\\
m_{2,2}\leq& \left\lfloor\log  \left(1+6P_{2,2}|h_2|^2\right)\right\rfloor,\label{con3}
\end{align}
where the flooring operation $\lfloor .\rfloor$ applies because the modulation orders must be integers. The motivation for introducing the modulation order constraints is to strike a balance between the achievable rate and the interference statistics. This can be seen by noting that under TIN and fixed channel gains, increasing the modulation order of only one user increases its achievable rate (until it reaches capacity) but also introduces more interference to other users. One can see that the RHS of \eqref{con1} is reminiscent of the single-user capacity of the strong user while the RHS of \eqref{con2} is reminiscent of the single-user capacity of user 2. The reason for excluding 1 inside the logarithm of \eqref{con2} while including 6 inside all logarithms is closely related to the minimum distance of individual constellation, which will be explained in Section \ref{sec:2up}c. By looking at \eqref{con1}, it is also worth noting that the sum capacity of the $K$-user downlink BC can be upper bounded by the single-user capacity of the strongest user \cite[Ch. 6.2.2]{tse_book}. Thus, one can regard \eqref{con1} as a sum-rate constraint. In addition, \eqref{con1} with $m_{2,1}=0$, \eqref{con2}, and \eqref{con3} are the individual modulation order constraints, where similar arguments apply.

\subsubsection{Power Assignments}\label{sec:2up}
We introduce two \emph{layers} power assignments. The first one is to assign power across different users' modulated symbols at the same time instant within the same sub-block of $\boldsymbol{x}$. The second layer power assignment is performed on top of the first layer power assignment by assigning power across different sub-blocks of $\boldsymbol{x}$.

$3a)$ \emph{First Layer Power Assignment}: The power to $v_1[j]$ and $v_2[j]$ for $j=1,\ldots,N_1$ is chosen such that the superimposed symbol satisfies
\begin{align}
v_1[j]+\sqrt{2^{m_1}}v_2[j]\in\Lambda_1+\sqrt{2^{m_1}}\Lambda_{2,1}, \label{eq:first_P_assign}
\end{align}
where the superimposed constellation $\Lambda_1+\sqrt{2^{m_1}}\Lambda_{2,1}$ is a regular QAM with cardinality $2^{m_1+m_{2,1}}$, zero mean, and minimum distance $d_{\min}(\Lambda_1+\sqrt{2^{m_1}}\Lambda_{2,1})=1$.

$3b)$ \emph{Second Layer Power Assignment}: On top of the first layer power assignment, we assign the power $P_1+P_{2,1}$ and $P_{2,2}$ to the first $N_1$ and the last $N_2-N_1$ symbols, respectively, of $\boldsymbol{x}$ such that the total power constraint is fulfilled. As a result, the transmitted signals for users 1 and 2 after the proposed two layers power assignments are
\begin{align}
x_1[j]=&\eta_1\sqrt{P_1+P_{2,1}}v_1[j],\; j=1,\ldots,N_1, \label{eq:x1} \\
x_2[j]=& \left\{ {\begin{array}{*{20}{c}}
\eta_1 \sqrt{2^{m_1}(P_1+P_{2,1})}v_2[j],\;j=1,\ldots,N_1\\
\eta_2\sqrt{P_{2,2}}v_2[j], \;j=N_1+1,\ldots,N_2\\
\end{array}} \right.,\label{eq:x2}
\end{align}
respectively, where $\eta_1 = \sqrt{\frac{6}{2^{m_1+m_{2,1}}-1}}$ and $\eta_2 = \sqrt{\frac{6}{2^{m_{2,2}}-1}}$ are the normalization factors to ensure both $\Lambda_1+\sqrt{2^{m_1}}\Lambda_{2,1}$ and $\Lambda_{2,2}$ have unit energy. For $X_1\overset{\text{unif}}{\sim}\eta_1\sqrt{P_1+P_{2,1}}\Lambda_1$ and $X_{2,1}\overset{\text{unif}}{\sim}\eta_1\sqrt{2^{m_1}(P_1+P_{2,1})}\Lambda_{2,1}$, and the power assignment in Section \ref{sec:2up}a, we obtain $P_1\hspace{-1mm}=\hspace{-1mm}\E[|X_1|^2]$ and $P_{2,1}\hspace{-1mm}=\hspace{-1mm}\E[|X_{2,1}|^2]$ where
\begin{align}
P_1 
=& \frac{2^{m_1}-1}{2^{m_1+m_{2,1}} -1}\left(\frac{N_2}{N_1}P - \frac{N_2-N_1}{N_1}P_{2,2}\right), \label{eq:P1}\\
P_{2,1} 
=&\frac{2^{m_1+m_{2,1}}-2^{m_1}}{2^{m_1+m_{2,1}} -1}\left(\frac{N_2}{N_1}P - \frac{N_2-N_1}{N_1}P_{2,2}\right).\label{eq:P21}
\end{align}
Thus, when $P_{2,2}$ is given, $P_1$ and $P_{2,1}$ become deterministic. In our scheme, we consider balanced second layer power assignment for each sub-block of superimposed symbol block $\boldsymbol{x}$ such that $P_{2,2}=P_1+P_{2,1}$. This also means that each sub-block of $\boldsymbol{v}_2$ has different power $P_{2,2}\neq P_{2,1}$ as long as $P_1 \neq 0$. We will show that this choice is good enough for the proposed scheme with QAM and TIN decoding to achieve rate pairs very close to those assume Gaussian and shell codes with perfect SIC and globally optimized $P_1$, $P_{2,1}$ and $P_{2,2}$ for maximizing achievable rate regions.


$3c)$ \emph{Minimum Distance}: By looking into the individual constellation while treating the other user's signals as noise, one can see that after the channel effects, i.e., $h_1x_1[j]\in h_1\eta_1\sqrt{P_1+P_{2,1}}\Lambda_1$ and $h_2x_2[j]\in h_2\eta_1\sqrt{P_1+P_{2,1}}\Lambda_{2,1}$, the minimum distance of each constellation satisfies
\begin{align}
d_{\min}\left(h_1\eta_1\sqrt{P_1+P_{2,1}}\Lambda_1\right)\overset{\eqref{con1}}{\geq}&1, \\
d_{\min}\left(h_2\eta_1 \sqrt{2^{m_1}(P_1+P_{2,1})}\Lambda_{2,1}\right)\overset{\eqref{con2}}{\geq}&1.\label{eq:dmin2}
\end{align}
Notice that in \eqref{eq:dmin2}, the logarithm in \eqref{con2} without 1 inside leads to a constant minimum distance lower bound. Hence, for any $(h_1,h_2)$ satisfying $|h_1|>|h_2|$, both constraints \eqref{con1}, \eqref{con2}, and the proposed power assignments in \eqref{eq:x1}-\eqref{eq:x2} \emph{guarantee constant minimum distance lower bound} for the superimposed constellation and each individual constellation after channel effects and normalization. The constant minimum distance lower bound is beneficial to TIN decoding for handling structural interference. It is worth noting that the structural interference comes from the fact that the interfering signal is uniformly distributed over a regular QAM in our design. In contrast, the conventional assumption of using Gaussian input distribution makes the interference Gaussian which is highly unstructured. As for $j=N_1+1,\ldots,N_2$, the constellation $\Lambda_{2,2}$ is already a regular QAM with $d_{\min}(\Lambda_{2,2})=1$ and $x_2[j]$ is interference-free. Thus, the first layer power assignment is not required here. With \eqref{con3}, one can easily verify that $d_{\min}(h_2\eta_2\sqrt{P_{2,2}}\Lambda_{2,2})\hspace{-1mm}\geq \hspace{-1mm}1$.



\subsubsection{TIN Decoding}
At the receiver, each user decodes its own messages by treating the other user's signals as noise. Hence, the other user's codebook information is completely unnecessary for the proposed scheme. For user $i$, $i\in\{1,2\}$, the decoder first computes the log-likelihood ratio (LLR) for each bit of the interleaved codeword $\boldsymbol{\tilde{c}}_i$ from the received signals $\boldsymbol{y}_i$ given in Section \ref{sec:model}. Then, the LLR sequence is deinterleaved and passed into a soft-input soft-out decoder. The decoding process is the same as that in the point-to-point channel using BICM \cite{CIT-019}.

\begin{remark}
When $|h_2|<|h_1|$, the first layer power assignment swaps the arguments between constellations $\Lambda_1$ and $\Lambda_{2,1}$ and their modulation orders in \eqref{eq:first_P_assign} in Section \ref{sec:2up}a. That is, we assign the power to $v_1[j]$ and $v_2[j]$ by $v_2[j]+2^{m_{2,1}}v_1[j]$ for $j=1,\ldots,N_1$. The rest of the steps do not change.
\demo
\end{remark}


%

\section{Finite Blocklength Achievable Rate Analysis}\label{sec:FBL}
In this section, we derive the second-order achievable rate of the downlink BC with discrete signaling and TIN with given blocklength and error probability constraints. The channel order is not required here due to the fact that TIN decoding is adopted. We define the normalized constellations after power assignments for $\boldsymbol{x}_1$ and the two sub-blocks of $\boldsymbol{x}_2$ as $\mathcal{X}_1$, $\mathcal{X}_{2,1}$, and $\mathcal{X}_{2,2}$, respectively.

\subsection{Achievable Rate of User 1}
We first analyze the information density, which is the key to our second-order achievable rate approximation. Based on the definition of information density in \cite{5452208}, the information density of user 1 is derived as
\begin{align}
&i(X_1^{[N_1]};Y_1^{[N_1]}) = \sum\nolimits_{j=1}^{N_1}i(X_1[j];Y_1[j])\label{eq:u1id_basic}\\
=&\hspace{-1mm}\sum_{j=1}^{N_1} \log\hspace{-1mm} \left( \frac{\sum\limits_{x_2[j] \in \mathcal{X}_{2,1}}P(y_1[j]|x_1[j],x_2[j])}{\sum\limits_{x_1[j] \in
\mathcal{X}_1}\sum\limits_{x_2[j] \in \mathcal{X}_{2,1}}P(y_1[j]|x_1[j],x_2[j])P(x_1[j])}\right)\hspace{-1mm}, \label{eq:u1id_basic1}
\end{align}
where we note that $X_1[j], X_2[j]$, and $Y_1[j]$ are i.i.d. for $j=1,\ldots,N_1$ and thus $i(X_1[j];Y_1[j])$ is also i.i.d., $P(y_i[j]|x_1[j],x_{2}[j])=\frac{1}{\pi}e^{-|y_i[j]-h_i(x_1[j]+x_{2}[j])|^2}$ for $i\in\{1,2\}$, and $P(x_1[j]) = \frac{1}{|\mathcal{X}_1|}$ and $P(x_2[j]) = \frac{1}{|\mathcal{X}_{2,1}|}$ due to uniform input distributions of $x_1[j]$ and $x_2[j]$, respectively, for $j=1,\ldots,N_1$. Then, we derive the mutual information for user 1 as
\begin{align}\label{eq:NI1}
I(X_1^{[N_1]};Y_1^{[N_1]})=N_1\E[i(X_1;Y_1)] = N_1I(X_1;Y_1),
\end{align}
where we have dropped the index $[j]$ because $i(X_1[j];Y_1[j])$ is i.i.d.. Further to \eqref{eq:NI1}, we derive $I(X_1;Y_1)$ in \eqref{eq:u1I}.
\begin{figure*}[t]
\begin{align}\label{eq:u1I}
I(X_1;Y_1)= \log|\mathcal{X}_1|-\frac{1}{|\mathcal{X}_{1}|\cdot|\mathcal{X}_{2,1}|}\sum_{x_{1}\in\mathcal{X}_{1}}\sum_{x_{2,1}\in\mathcal{X}_{2,1}}\E_{Z_1}\left[\log \left(\frac{\sum\limits_{x'_1\in \mathcal{X}_1}\sum\limits_{x'_{2,1}\in \mathcal{X}_{2,1}}e^{-|Z_1+h_1(x_1-x'_1+x_{2,1}-x'_{2,1})|^2}}{\sum\limits_{x'_{2,1}\in \mathcal{X}_{2,1}}e^{-|Z_1+h_1(x_{2,1}-x'_{2,1})|^2}}\right) \right].
\end{align}
\hrule
\end{figure*}
Next, we derive the dispersion function for user 1 as
\begin{align}
V(X_1^{[N_1]};&Y_1^{[N_1]}) 
\overset{\eqref{eq:u1id_basic}}{=}\text{Var}\left[\sum\nolimits_{j=1}^{N_1}i(X_1[j];Y_1[j])\right]\\
=&\sum\nolimits_{j=1}^{N_1}\text{Var}[i(X_1[j];Y_1[j])]
= N_1V(X_1;Y_1),\label{eq:u1Vint}
\end{align}
where \eqref{eq:u1Vint} holds because $x_1[j]$ and $x_1[j']$ are independent and $y_1[j]$ and $y_1[j']$ are independent for any $j\neq j'$ and $j,j'\in \{1,\ldots,N_1\}$. We then derive $V(X_1;Y_1)$ in \eqref{eq:u1V}.
\begin{figure*}[t]
\begin{align}
V(X_1;Y_1)
=&\frac{1}{|\mathcal{X}_{1}|\cdot|\mathcal{X}_{2,1}|}\sum_{x_{1}\in\mathcal{X}_{1}}\sum_{x_{2,1}\in\mathcal{X}_{2,1}}\E_{Z_1}\left[\left(\log\left( \frac{\sum\limits_{x'_1\in \mathcal{X}_1}\sum\limits_{x'_{2,1}\in \mathcal{X}_{2,1}}e^{-|Z_1+h_1(x_1-x'_1+x_{2,1}-x'_{2,1})|^2}}{\sum\limits_{x'_{2,1}\in \mathcal{X}_{2,1}}e^{-|Z_1+h_1(x_{2,1}-x'_{2,1})|^2}}\right)\right)^2 \right]\nonumber \\
&-\left(\frac{1}{|\mathcal{X}_{1}|\cdot|\mathcal{X}_{2,1}|}\sum_{x_{1}\in\mathcal{X}_{1}}\sum_{x_{2,1}\in\mathcal{X}_{2,1}}\E_{Z_1}\left[\log \left(\frac{\sum\limits_{x'_1\in \mathcal{X}_1}\sum\limits_{x'_{2,1}\in \mathcal{X}_{2,1}}e^{-|Z_1+h_1(x_1-x'_1+x_{2,1}-x'_{2,1})|^2}}{\sum\limits_{x'_{2,1}\in \mathcal{X}_{2,1}}e^{-|Z_1+h_1(x_{2,1}-x'_{2,1})|^2}}\right) \right]\right)^2 \label{eq:u1V}.
\end{align}
\hrule
\end{figure*}

We then have the following proposition for the second-order achievable rate of user 1.
\begin{proposition}\label{prof:u1}
Define $\epsilon_1$ to be the upper bound on the average TIN decoding error probability of user 1. For the channel model in \eqref{eq:y1}, user 1's achievable rate by treating user 2's signals as interference is bounded by
\begin{align}\label{eq:u1rate}
R_1 \hspace{-1mm} \leq \hspace{-1mm} I(X_1;Y_1)\hspace{-1mm}-\hspace{-1mm}\sqrt{\frac{V(X_1;Y_1)}{N_1}}Q^{-1}\left(\epsilon_1\right)\hspace{-1mm}+\hspace{-1mm}O\left(\frac{\log  N_1}{N_1}\right)\hspace{-1mm}, \hspace{-1mm}
\end{align}
where $I(X_1;Y_1)$ is in \eqref{eq:u1I} and $V(X_1;Y_1)$ is in \eqref{eq:u1V}.
\end{proposition}
\emph{Proof of Proposition \ref{prof:u1} (Sketch):} We denote by $M_1$ the codebook size for user 1. First, user 1's decoding error probability under TIN as a function of $N_1$ can be upper bounded by using the dependence testing bound \cite[Th. 17]{5452208}
\begin{align}
\epsilon_1(N_1) \leq& \E\left[2^{-\max\left\{0,i(X_1^{[N_1]};Y_1^{[N_1]})-\log \frac{M_1-1}{2}\right\}}\right]\\
\leq& \mathbb{P}\left[ \frac{M_1-1}{2}2^{-i(X^{[N_1]},Y^{[N_1]})}> \frac{1}{\sqrt{N_1}}\right]\hspace{-1mm}+\hspace{-1mm}\frac{1}{\sqrt{N_1}}.\label{eq:RCU}
\end{align}
With \eqref{eq:RCU}, we then use the Berry-Esseen central limit theorem \cite[Th. 2, Ch. XVI-5]{Feller_book} and get
\begin{align}\label{eq:CLT_u1}
\epsilon_1(N_1) \leq& Q\left( \frac{N_1I(X_1;Y_1)-\log  \frac{M_1-1}{2} -\log  \sqrt{N_1}}{\sqrt{N_1 V(X_1;Y_1)}}\right) \nonumber \\
&+O\left(\frac{1}{\sqrt{N_1}}\right) \leq \epsilon_1,
\end{align}
where we have used the properties in \eqref{eq:NI1} and \eqref{eq:u1V}. The last inequality of \eqref{eq:CLT_u1} ensures that the error probability \eqref{eq:RCU} is upper bounded by $\epsilon_1$ for all $N_1$. One can then solve for $\log (M_1-1)$ and perform the first-order Taylor expansion of $Q^{-1}(.)$ about $\epsilon_1$. Finally, dividing both sides of the resultant inequality by $N_1$ and using the fact that $R_1=\frac{\log M_1}{N_1}\leq\frac{\log (M_1-1)+1}{N_1}$ for $M_1\geq 2$, we obtain \eqref{eq:u1rate}. \demo

Since user 1 has the shortest symbol blocks, each intended symbol for user 1 experiences the same interference statistics, which is similar to the homogeneous blocklength case. However, the interference experienced by user 2 behaves differently from user 1 as we will see in the next section.


\subsection{Achievable Rate of User 2}\label{sec:u2}
Since $\boldsymbol{x}_2$ will be partially interfered, $X_2[j]$ and $Y_2[j]$ are i.i.d. when either $j = 1,\ldots,N_1$ or $j=N_1+1,\ldots,N_2$ whereas $X_2[j](Y_2[j])$ and $X_2[j'](Y_2[j'])$ are \emph{not necessarily} identically distributed for $j\in\{1,\ldots,N_1\}$ and $j'\in\{N_1+1,\ldots,N_2\}$. In this case, we let $X_{2,1}$($Y_{2,1}$) to represent the random variables $X_2[j]$($Y_2[j]$) for $j \in \{1, . . . ,N_1\}$ and let $X_{2,2}$($Y_{2,2}$) to represent the random variables $X_2[j]$($Y_2[j]$) for $j \in \{N_1 + 1, . . . ,N_2\}$. The information density for user 2 is
\begin{align}
&i(X_2^{[N_2]};Y_2^{[N_2]}) = \sum\nolimits_{j=1}^{N_2}i(X_2[j];Y_2[j]) \label{eq:u2id_basic}\\
&=\hspace{-1mm}\sum_{j=1}^{N_1}\log  \left( \frac{\sum\limits_{x_1[j]\in\mathcal{X}_1}P(y_2[j]|x_2[j],x_1[j])}{\sum\limits_{x_2[j]\in\mathcal{X}_{2,1}}\sum\limits_{x_1[j]\in\mathcal{X}_1}P(y_2[j]|x_2[j],x_1[j])P(x_2[j])}\right)\nonumber \\
&+\hspace{-2mm}\sum\limits_{j=N_1+1}^{N_2}\log \left( \frac{P(y_2[j]|x_2[j])}{\sum\limits_{x_2[j]\in\mathcal{X}_{2,2}}P(y_2[j]|x_2[j])P(x_2[j])} \right).\label{eq:u2id}
\end{align}
With \eqref{eq:u2id}, the mutual information under TIN for user 2 is
\begin{align}\label{eq:NI2}
I(X_2^{[N_2]};Y_2^{[N_2]})=\sum\nolimits^2_{i=1}(N_i-N_{i-1})I(X_{2,i};Y_{2,i}),
\end{align}
where $I(X_{2,1};Y_{2,1})$ can be easily obtained similarly to \eqref{eq:u1I} by swapping the arguments between user 1 and user 2, and $I(X_{2,2};Y_{2,2})$ is the mutual information of the single-user channel. Next, we derive the dispersion function as
\begin{align}\label{eq:N2V2}
V(X_2^{[N_2]};Y_2^{[N_2]}) =\sum\nolimits^2_{i=1}(N_i-N_{i-1})V(X_{2,i};Y_{2,i}),
\end{align}
where $V(X_{2,1};Y_{2,1})$ can be easily obtained from \eqref{eq:u1V} by swapping the arguments between user 1 and user 2, and $V(X_{2,2};Y_{2,2})$ is the dispersion of the single-user channel.

Having derived the mutual information and dispersion, we have the following proposition for the second-order achievable rate of user 2.
\begin{proposition}\label{prof:u2}
Define $\epsilon_2$ to be the upper bound on the average TIN decoding error probability of user 2. For the channel model in \eqref{eq:y2}, user 2's achievable rate by treating user 1's signals as interference is bounded by
\begin{align}\label{eq:u2rate}
R_2 \leq& \frac{\sum^2_{i=1}N_i-N_{i-1}}{N_2}I(X_{2,i};Y_{2,i}) \nonumber\\
&-\frac{\sqrt{\sum^2_{i=1}(N_i-N_{i-1})V(X_{2,i};Y_{2,i})}}{N_2}Q^{-1}\left(\epsilon_2\right)\nonumber\\
&+O\left(\frac{\log  N_2}{N_2}\right).
\end{align}
\end{proposition}
The proof of Proposition \ref{prof:u2} follows from that of Proposition \ref{prof:u1} and is omitted due to space limitation. The impacts of the length of interfering symbols on user 2's achievable rate are clearly shown in \eqref{eq:u2rate}. This is different from the homogeneous blocklength case for which a single signal-to-interference-plus-noise ratio (SINR), i.e., $\frac{P_2|h_2|^2}{P_1|h_2|^2+1}$, could not capture the effects of partially interfered symbol sequences.

\subsection{Modulation and Code Design}
With the derived achievable rates, we can design the modulations and channel codes for the proposed schemes. First, consider the blocklength $(N_1,N_2)$ and error probability $(\epsilon_1,\epsilon_2)$ requirements for both users. We design the modulations $(\Lambda_1,\Lambda_{2,1},\Lambda_{2,2})$ whose orders satisfying \eqref{con1} such that the achievable rate pair computed by using \eqref{eq:u1rate} and \eqref{eq:u2rate} reach a target rate pair $(R_1,R_2)$. Note that given the modulations, the power assignments become deterministic according to Section \ref{sec:2up}. Moreover, users 1 and 2's codeword lengths satisfy $(n_1,n_2)=(N_1m_1,N_1m_{2,1}+(N_2-N_1)m_{2,2})$ according to \eqref{u1n1} and \eqref{u2n2} in Section \ref{sec:2u_const}. To match users 1 and 2's transmission rates with their corresponding achievable rates, i.e., $(R_1,R_2) =  (\frac{k_1}{n_1}m_1,\frac{k_2}{n_2}(\frac{N_1}{N_2}m_{2,1}+\frac{N_2-N_1}{N_2}m_{2,2}))$, the information lengths of users 1 and 2's channel codes are obtained as $(k_1,k_2)=(R_1N_1,R_2N_2)$. The problem can now be converted into designing good point-to-point codes with the specified information and codeword lengths.

\section{Simulation Results}\label{sec:sim}
\subsection{Achievable Rate}\label{sec:rate_sim}
We present a design example to showcase the performance of the proposed scheme with QAM and TIN. We define the signal-to-noise ratio $\SNR_i \triangleq P|h_i|^2$ for user $i\in\{1,2\}$. We consider $(\SNR_1,\SNR_2)=(18,5)$ in dB, $(N_1,N_2)=(128,256)$, and $(\epsilon_1,\epsilon_2)=(10^{-6},10^{-4})$. For comparison purposes, we have included two benchmark schemes using shell codes and Gaussian codes and assume perfect SIC regardless of blocklength. We stress that the perfect SIC assumption in the benchmark schemes is used for comparison purposes only. Hence, the performance of the benchmark schemes with perfect SIC serves as an upper bound of all achievable schemes that take into account imperfect SIC, e.g., \cite{9518265,9838392}. The achievable rates of the benchmark schemes can be derived by following from Section \ref{sec:FBL}. Note that although both Gaussian and shell codes achieve capacity in the infinite blocklength regime, shell codes have a larger second-order achievable rate due to smaller dispersion \cite[Eqs. (23)\&(25)]{7605463}.

\begin{figure}[t!]
	\centering
\includegraphics[width=3.4in,clip,keepaspectratio]{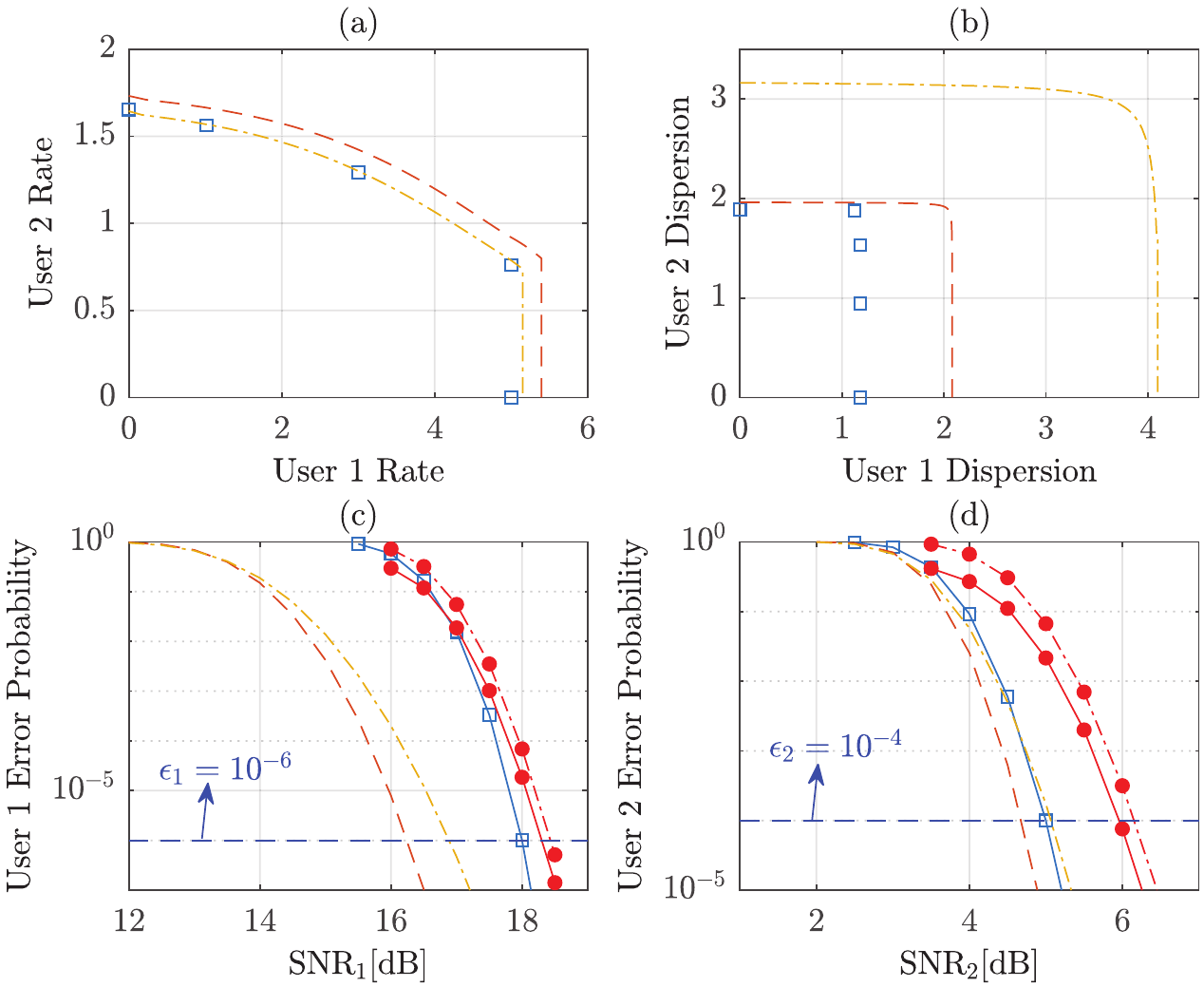}
\caption{(a) Achievable rate; (b) Dispersion; (c) Error probability of user 1; (d) Error probability of user 2. Label: $\square$ QAM without SIC analytical, $--$ Shell codes with perfect SIC, $-\cdot-\cdot$ Gaussian codes with perfect SIC, $-\hspace{-1mm}\bullet\hspace{-1mm}-$ 5G CA-polar BER, $-\cdot\bullet-\cdot$ 5G CA-polar BLER.}
\label{fig:ar2}
\end{figure}



The second-order achievable rate pairs (without the third-order term) and the corresponding dispersion of the proposed scheme and the aforementioned two benchmark schemes are shown in Figs. \ref{fig:ar2}(a) and \ref{fig:ar2}(b), respectively. The modulation orders for the proposed scheme are $(m_1,m_{2,1},m_{2,2})=(0,4,4),(2,4,4),(4,2,4),(6,0,4),(6,0,0)$, corresponding to the data points from left to right in the figure. Note that the proposed scheme uses balanced second layer power assignment, i.e., $P_1+P_{2,1}=P_{2,2}$ while the two benchmark schemes use brute-force search for $(P_1,P_{2,1},P_{2,2})$ to obtain their largest possible rate regions. Observe that the proposed scheme with QAM and TIN can achieve rate pairs very close to those under Gaussian signaling and perfect SIC. Meanwhile it is shown in Fig. \ref{fig:ar2}(b) that the dispersion of QAM in the proposed scheme is much smaller than that of Gaussian codes and is no larger than that of shell codes. Since short blocklength and ultra-low target error probability are the main features of URLLC communication scenarios, the second-order term has a substantial impact on the achievable rate. Hence, our results demonstrate that the proposed scheme is promising in supporting heterogeneous URLLC services. Notice that user 1's rate can remain to be the single-user rate while user 2's rate is increasing as shown in the bottom right corner of Fig. \ref{fig:ar2}(a). This is achieved by setting $P_{2,1}=0$ such that increasing $P_{2,2}$ does not affect user 1's second-order achievable rate.

\subsection{Error Probability}\label{sec:err}

We build a practical set-up of the proposed scheme by using off-the-shelf codes. For illustrative purpose, we consider the same channel setting as in Section \ref{sec:rate_sim}, where the proposed scheme with modulation orders $(m_1,m_{2,1},m_{2,2})=(2,4,4)$ achieves a rate pair of $(R_1,R_2) = (1.0174,1.5644)$. Since $(N_1,N_2)=(128,256)$, the channel codes for users 1 and 2 are with $(n_1,k_1)=(256,130)$ and $(n_2,k_2)=(1024,400)$, respectively. Each user employs a 5G standard CRC-aided polar (CA-polar) code with an 11-bit CRC and adopts successive-cancellation list decoding \cite{TS138212_v16p8}. We set the decoding list size 32 for user 1 and 64 for user 2. The bit error rate (BER) and block error rate (BLER) for users 1 and 2 are reported in Figs. \ref{fig:ar2}(c) and \ref{fig:ar2}(d), respectively. We also include the average block error probability upper bound of the benchmark schemes with Gaussian codes and shell codes with perfect SIC as well as that of the proposed scheme with QAM and TIN. The error probability is obtained by rearranging the second-order achievable rate, i.e., \eqref{eq:u1rate} and \eqref{eq:u2rate}. Note that all schemes achieve the same rate pair and use the same power allocation.


Observe that the BER and BLER of user 1 are more close to the analytical error probability of QAM at $10^{-6}$ than that for user 2 at $10^{-4}$. In fact, this behavior is similar to the single-user case \cite{8594709}, where CA-polar codes with short blocklength perform better than that with moderate blocklength. This implies that the proposed scheme can allow the good performance of a code on the point-to-point AWGN channel to be \emph{carried over} to the considered multiuser channel under heterogeneous interference. It is also interesting to see that for user 2, the error probability upper bound for QAM slightly outperforms that of Gaussian signaling at $10^{-4}$ and below. This demonstrates that the proposed scheme with the simplest TIN decoding is very promising at short blocklength. For user 1, the analytical error probability of QAM is about 1 dB away from that of the Gaussian code at $10^{-6}$. This is because user 1 does not perform SIC in the proposed scheme while the benchmark schemes assume perfect SIC. Hence, the error performance of the corresponding coded systems also shows similar behavior. In summary, the proposed scheme using off-the-shelf codes can achieve satisfactory performance when compared to the error performance of shell codes with perfect SIC assumption. 



\section{Conclusion}
In this paper, we have proposed a new transmission scheme based on discrete signaling and TIN for the downlink BC under heterogeneous blocklength and error probability constraints. To effectively handle heterogeneous interference across received symbol sequences, we have divided the symbol block of each user into sub-blocks and designed the modulation and power for each sub-block. We also have derived the second-order achievable rate under practical modulations and TIN to characterize the behavior of practical coded modulation systems for the considered scenario. Simulation results have shown that under short blocklength constraints, the proposed scheme with QAM and TIN can operate very close to the benchmark schemes that assume perfect SIC with Gaussian signaling. This implies that practical coded modulations together with the simplest single-user decoding are very promising for supporting downlink multiplexing of heterogeneous services with desired latency and reliability requirements while achieving near-optimal rates.

%
%


\bibliographystyle{IEEEtran}
\bibliography{MinQiu}

\end{document}